\documentclass{PoS}

\def\a{{\alpha}}
\def\b{{\beta}}
\def\d{{\delta}}
\def\e{{\epsilon}}
\def\g{{\gamma}}
\def\k{{\kappa}}
\def\s{{\sigma}}
\def\t{{\tau}}
\def\tel{{\tau_{el}}}
\def\L{{\Lambda}}

\def\mc#1{{\mathcal #1}}

\title{Cottingham formula for the electromagnetic self-energy contribution to $M_p - M_n$}

\ShortTitle{Cottingham Formula}

\author{\speaker{Andr\'{e} Walker-Loud}\\
        Lawrence Berkeley National Laboratory, Berkeley, CA 94720, USA\\
        Department of Physics, University of California, Berkeley, CA 94720, USA \\
        E-mail: \email{awalker-loud@lbl.gov}}

\author{Carl E. Carlson\\
        Department of Physics, College of William and Mary, Williamsburg, VA 23187-8795.\\
        E-mail: \email{carlson@physics.wm.edu}}
\author{Gerald A. Miller\\
        Department of Physics, University of Washington, Seattle, WA 98195-1560.\\
        E-mail: \email{miller@phys.washington.edu}}

\abstract{
We provide an update of the determination of the electromagnetic self-energy contribution to $M_p - M_n$ based upon Cottingham's Formula.
A technical oversight in the literature is uncovered: 
the application of the Cottingham Formula requires the use of a subtracted dispersion integral;
an argument to evade the subtraction function was presented;
the argument was based on false assumptions about the scaling violations of the parton model, a point first mentioned by J.~C. Collins.
We elucidate this point and utilize low-energy effective theory to relate the unknown subtraction function to the nucleon isovector magnetic polarizability.
This allows us to provide the first reliable determination of $\d M^\g = 1.30(03)(47)$~MeV~\cite{WalkerLoud:2012bg}.
}

\FullConference{The 30 International Symposium on Lattice Field Theory - Lattice 2012,\\
		June 24-29, 2012\\
		Cairns, Australia}

\begin{document}

%
%%		INTRODUCTION
%%%
%%%%
\section{Introduction}

Given only electrostatic forces, a natural prediction is $M_p > M_n$ yet in nature~\cite{MPmnsMN}
\begin{equation}\label{eq:MPmnsMN}
M_p - M_n = -1.29333217(42) \textrm{ MeV}\, .
\end{equation}
Before we knew of QCD, there were many attempts to reconcile this apparent discrepancy, see Ref.~\cite{Zee:1971df} for a review.
We now know the Standard Model has two sources of isospin breaking: the electromagnetic couplings of the light quarks and their mass parameters.
The contributions to the nucleon mass splitting from these two sources are roughly equal in magnitude but opposite in sign.
While the net result is well known, Eq.~(\ref{eq:MPmnsMN}), our ability to disentangle the contribution from these two sources remains relatively poorly constrained.

%The nucleon mass splitting plays a critical role in the composition of matter in the early universe.  We know the abundance of light nuclei after Big Bang Nucleosynthesis (BBN) was roughly 75\% H and 25\% ${}^4$He by mass fraction which means there was roughly one neutron for every 7 protons.
%Understanding the primordial neutron to proton ratio ($X_n/X_p$) depends on light nuclear reactions, but the initial conditions for BBN are simply given by
%\begin{equation}
% X_n / X_p = e^{-(M_n - M_p) / T}\, ,
%\end{equation}
%for our universe at temperature $T$ (after the neutrinos decouple from thermodynamic equilibrium, roughly $0.1$~seconds after the Big Bang when $T\sim3$~MeV~$\sim 3\times10^{10}$~K).
%In fact, our understanding of BBN is sufficiently quantitative as to provide stringent constraints on proposed variations of the fundamental constants, see Ref.~\cite{BBN} for a review.

An understanding of Eq.~(\ref{eq:MPmnsMN}) from first principles is desired.
For example, a determination of the electromagnetic contribution to the nucleon mass splitting will allow for an independent means of determining $m_d - m_u$~\cite{WalkerLoud:2010qq}.
At leading order (LO) in isospin breaking, we can write
\begin{equation}
\d M^{p-n} = M_p - M_n = \d M^\g + \d M^{m_d - m_u}\, .
\end{equation}
The quark mass operator is needed to renormalize electromagnetic self-energy contribution, entangling these effects beyond LO.
In general, electromagnetic effects can not be unambiguously separated from hadronic effects so one must formulate a prescription to disentangle them~\cite{Gasser:2003hk}.
At LO, lattice QCD can be used to reliably compute the contribution from $m_d - m_u$.  A weighted, uncorrelated average from 3 independent lattice calculations~\cite{Beane:2006fk} yields,%
%FOOTNOTE
\footnote{There is an additional determination in Ref.~\cite{Horsley:2012fw} which we have not yet included.}
\begin{equation}
	\d M^{m_d - m_u}_{LQCD} = -2.53(40) \textrm{ MeV}\, .
\end{equation}
Calculating $\d M^\g$ with lattice QCD and QED is much more challenging, due to the disparate length scales relevant for the two theories.
There is an alternative method for computing $\d M^\g$ known as the Cottingham formula~\cite{Cottingham:1963zz}, which relates the electromagnetic self-energy to an integral of the forward Compton scattering tensor.
In 1975, Gasser and Leutwyler provided the determination~\cite{Gasser:1974wd}
\begin{equation}
\d M^\g = 0.76(30) \textrm{ MeV}\, .
\end{equation}
If we instead subtract the lattice QCD determination from experiment, we arrive at
\begin{equation}
\d M^{p-n} - \d M^{m_d - m_u}_{LQCD} = 1.24(40) \textrm{ MeV}\, .
\end{equation}
While the central values in these two determinations are quite different, they are only one-sigma discrepant.
A more precise determination of either $\d M^\g$ or $\d M^{m_d - m_u}$ is clearly desirable.
We report on an effort to improve the electromagnetic self-energy contribution using the Cottingham Formula~\cite{WalkerLoud:2012bg}.

%
%%		Cottingham
%%%
%%%%
\section{The Cottingham Formula}
%%%%%%%%%%%%%%%%%%%%%%%%%%%%%%%%%%%%%%%%%%%%%%
\begin{figure}
\center
\begin{tabular}{cc}
\includegraphics[width=0.35\textwidth]{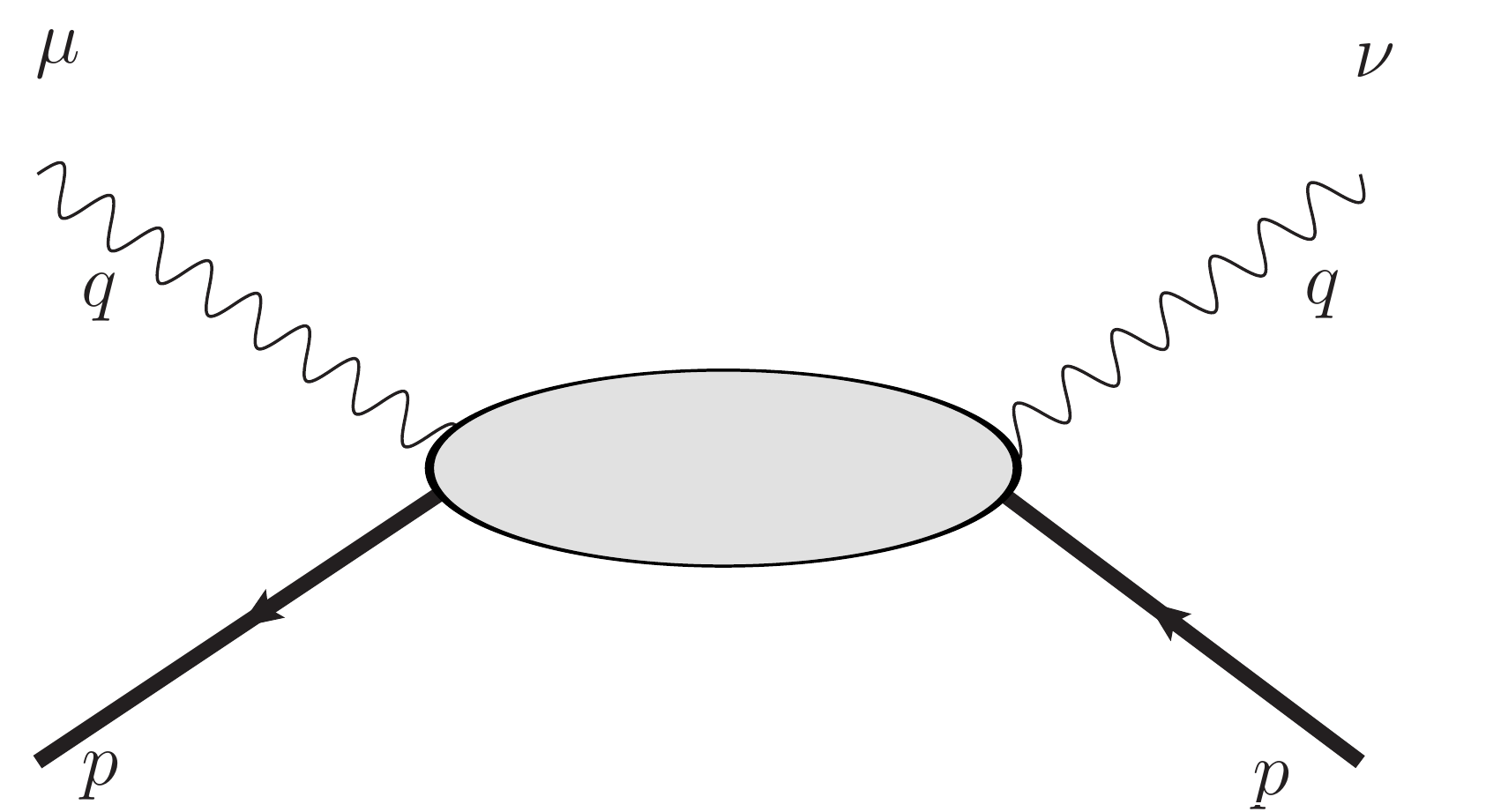}
&
\includegraphics[width=0.35\textwidth]{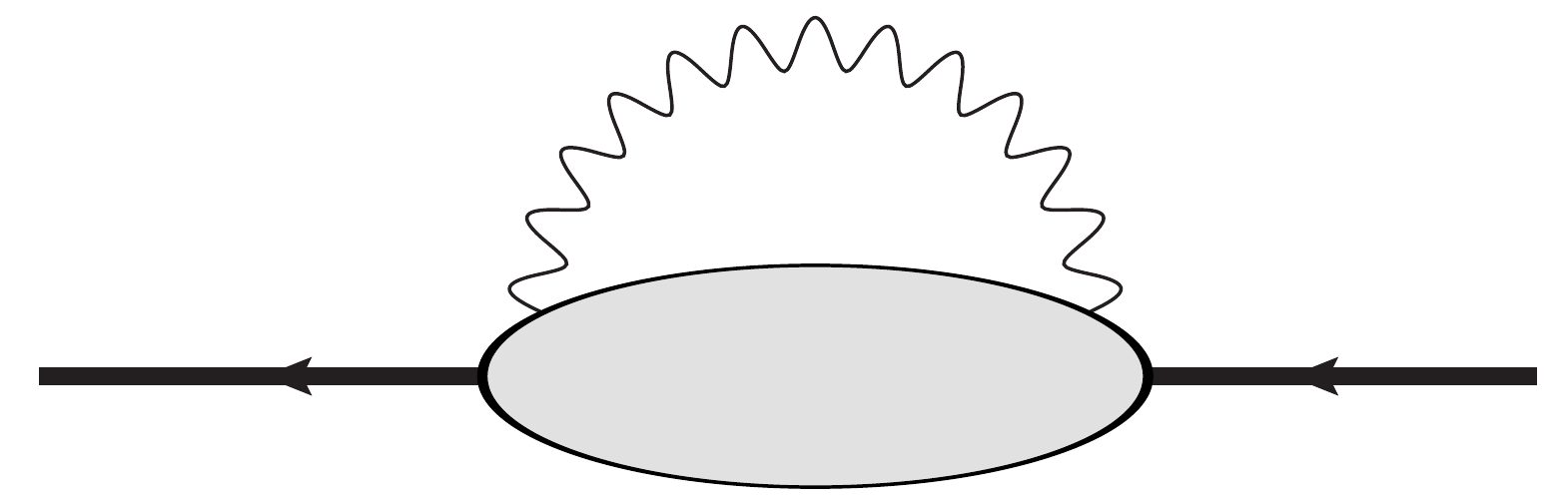}
\\ (a) & (b)
\end{tabular}
\caption{\label{fig:Cottingham}  The Cottingham Formula intuitively relates the forward Compton Scattering (a) to the $\mathcal{O}(\alpha_{f.s.})$ electromagnetic self-energy (b).}
\end{figure}
%%%%%%%%%%%%%%%%%%%%%%%%%%%%%%%%%%%%%%%%%%%%%%

\begin{figure}
\center\includegraphics[width=0.5\textwidth]{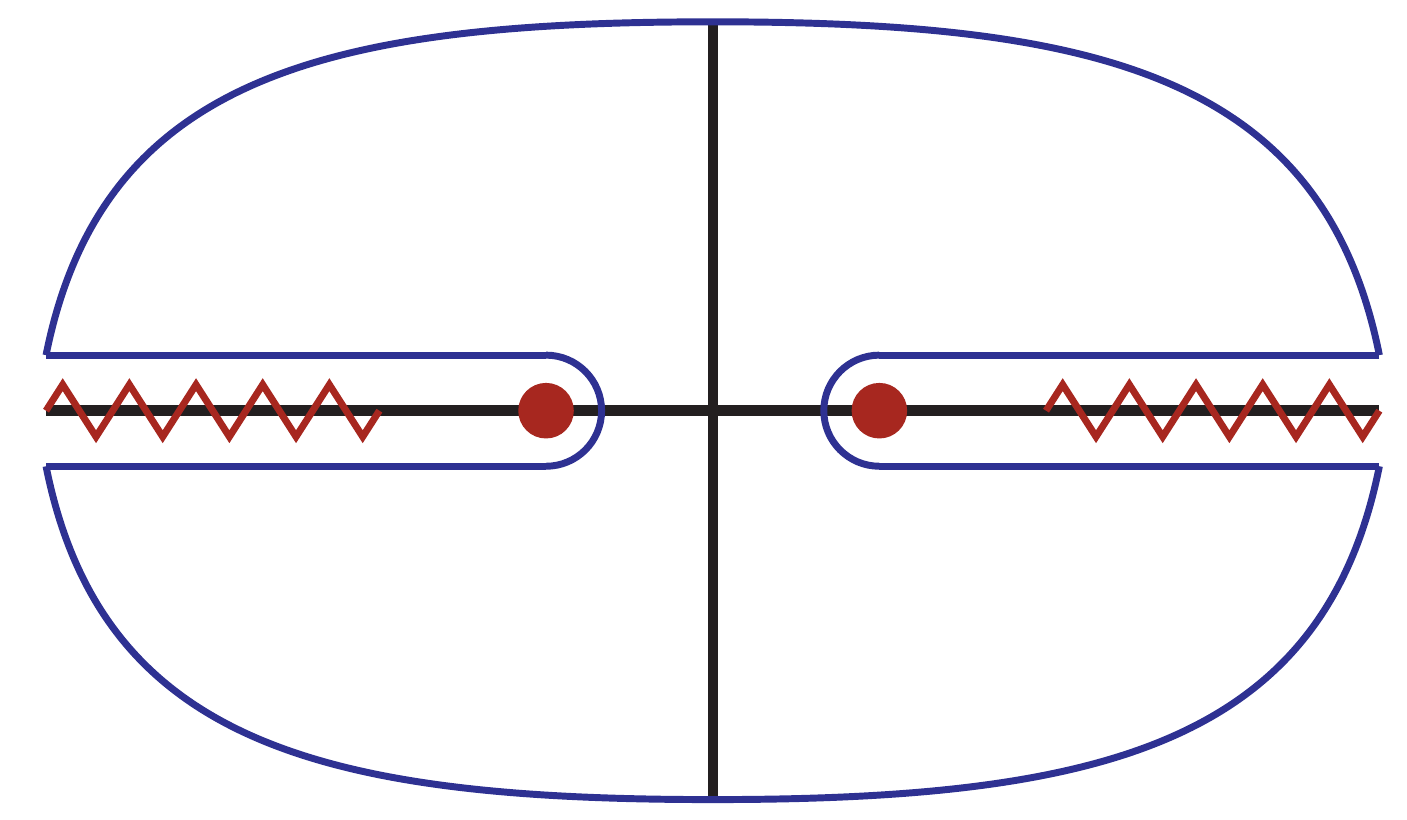}
\caption{\label{fig:dispersion}  Analytic structure of scalar functions $T_i(\nu,Q^2)$ in the complex $\nu$ plane for fixed $Q^2$, and the contour used to evaluate them.}
\end{figure}

The Cottingham Formula intuitively relates the electromagnetic self-energy to  forward Compton Scattering (see Fig.~\ref{fig:Cottingham})
\begin{equation}\label{eq:dMg}
\d M^\g = \frac{i}{2M} \frac{\alpha_{f.s.}}{(2\pi)^3}
	\int_R d^4 q \frac{g^{\mu\nu}}{q^2 +i\e} T_{\mu\nu}(p,q)\, .
\end{equation}
The subscript $R$ on the integral reminds us that the self-energy contains a logarithmic divergence and must be renormalized~\cite{Collins:1978hi}.  The spin-averaged forward Compton Scattering tensor is given by
\begin{eqnarray}\label{eq:Tmunu}
T_{\mu\nu}(p,q) &=& \frac{i}{2}\sum_\sigma \int d^4 x e^{iq\cdot x} 
	\langle p\sigma | T\left\{ J_\mu(x) J_\nu(0) \right\} | p\sigma \rangle
\nonumber\\
	&=&
	\left(\frac{q_\mu q_\nu}{q^2} -g_{\mu\nu} \right) T_1(q^0,-q^2)
	+\left(p_\mu - q_\mu \frac{p\cdot q}{q^2} \right) \left(p_\nu - q_\nu \frac{p\cdot q}{q^2} \right) 
		\frac{T_2(q^0,-q^2)}{M^2}
\end{eqnarray}
Cottingham showed that by performing the Wick rotation $q^0 \rightarrow i\nu$ and then a variable transformation $Q^2 = \mathbf{q}^2 +\nu^2$, the nucleon self-energy can be related to the experimentally measured structure functions;
\begin{equation}\label{eq:Cottingham}
\d M^\g = \frac{\alpha_{f.s.}}{8\pi^2} \int_0^{\L^2} {\hskip-0.8em} dQ^2 \int_{-Q}^{+Q} {\hskip-0.8em}d\nu
	\frac{\sqrt{Q^2 - \nu^2}}{Q^2} \frac{T_\mu^\mu}{M}
	+\d M^{c.t.}(\L)\, ,
\end{equation}
where
\begin{equation}
T_\mu^\mu = -3 T_1(i\nu,Q^2) + \left(1 - \frac{\nu^2}{Q^2} \right) T_2(i\nu,Q^2)\, .
\end{equation}
One uses fixed-$Q^2$ dispersion integrals to determine the scalar functions $T_i(i\nu,Q^2)$ in terms of their experimentally measured absorptive (imaginary) parts.
It is known that $T_2$ satisfies an unsubtracted dispersion integral while $T_1$ requires one subtraction~\cite{Harari:1966mu}.
These scalar functions are crossing symmetric $T_i(-\nu,Q^2) = T_i(\nu,Q^2)$, thus given by
\begin{eqnarray}
T_1(\nu,Q^2) &=& T_1(0,Q^2) + \frac{\nu^2}{2\pi} \int_{\nu_{th}}^\infty {\hskip-0.8em} d\nu^\prime
	\frac{2\nu^\prime }{(\nu^\prime)^2 ( (\nu^\prime)^2 - \nu^2)} 2\textrm{Im}\, T_1(\nu^\prime+i\e,Q^2)
\\
T_2(\nu,Q^2) &=& \frac{1}{2\pi} \int_{\nu_{th}}^\infty {\hskip-0.8em} d\nu^\prime
	\frac{2\nu^\prime }{(\nu^\prime)^2 - \nu^2} 2\textrm{Im}\, T_2(\nu^\prime+i\e,Q^2)
\end{eqnarray}
The $+i\e$ in the argument indicates the function is evaluated just above the cut on the positive real axis (see Fig.~\ref{fig:dispersion}).  
The absorptive parts are given in terms of the well known nucleon structure functions 
%%%%%%%%%%%%%%%%%%%%%%%%%%%%%%%%%%%%%%%%%%%%%%
%%%%%%%%%%%%%%%%%%%%%%%%%%%%%%%%%%%%%%%%%%%%%%
\begin{equation}
	2\textrm{Im}\, T_1(\nu,Q^2) = 2\pi\ F_1(\nu,Q^2)\, ,
	\qquad
	2\textrm{Im}\, T_2(\nu,Q^2) = 2\pi\ \frac{M}{\nu}F_2(\nu,Q^2)\, ,
\end{equation}
to which we have implicitly included the isolated elastic nucleon pole.%
%FOOTNOTE
\footnote{The elastic pole is isolated because we are working to leading order in QED, so there is no Bremsstrahlung radiation.  The inelastic cut begins at the pion production threshold.} 
Except in the low and high $Q^2$ limits, the subtraction function $T_1(0,Q^2)$ can not be simply related to measured cross sections, complicating the determination of $\d M^\g$.

The presence of the unknown subtraction function severely impacts the ability to determine $\d M^\g$ precisely.  In Ref.~\cite{Gasser:1974wd}, an argument to evade the subtraction function based on the parton model was presented.
However, as was first noted in Ref.~\cite{Collins:1978hi}, the argument was based on false assumptions about the scaling violations of the Callan-Gross relation~\cite{Callan:1969uq}.
To understand the issue, we first recall that the decomposition of the Compton tensor, Eq.~(\ref{eq:Tmunu}) is not unique.
Another common decomposition is
\begin{eqnarray}\label{eq:tmunu}
T_{\mu\nu}(p,q) &=&
	\left(g_{\mu\nu} - \frac{q_\mu q_\nu}{q^2} \right) q^2 t_1(q^0,-q^2)
\nonumber\\ &&
	-\left[ p_\mu p_\nu 
		-\frac{p\cdot q}{q^2} \left( p_\mu q_\nu +p_\nu q_\mu \right) 
		+\frac{(p\cdot q)^2}{q^2} g_{\mu\nu} 
	\right]
	\frac{q^2 t_2(q^0,-q^2)}{M^2}\, .
\end{eqnarray}
What are the advantages of this choice over Eq.~(\ref{eq:Tmunu})?
For a point particle, such as the electron, at leading order in QED, $t_1 = 0$.
In general
\begin{equation}\label{eq:t1}
	2\textrm{Im}\, t_1(\nu,Q^2) = \frac{2\pi M \nu}{Q^4} \left[ 2x F_1(x,Q^2) - F_2(x,Q^2) \right]
	\quad \textrm{where}
	\quad x = \frac{Q^2}{2M\nu}\, .
\end{equation}
In the deep-inelastic (DIS) limit $2xF_1(x) - F_2(x) = 0$.
In Ref.~\cite{Gasser:1974wd}, it was assumed the corrections to this Callan-Gross relation were power law, 
\begin{equation}
	2xF_1 - F_2 \rightarrow \frac{H_1(x)}{\nu}\, ,
\end{equation}
as predicted by the parton model.  This extra suppression in $\nu$ would allow for an evasion of the unknown subtraction function.  However, it is known in QCD that the corrections are only suppressed by the strong coupling constant~\cite{Zee:1974du};
\begin{equation}
	2xF_1(x,Q^2) - F_2(x,Q^2) = -\frac{32}{9} \frac{\a_S(Q^2)}{4\pi} F_2(x,Q^2)
\end{equation}
In the Regge limit (fixed $Q^2$, $\nu \rightarrow \infty, x\rightarrow 0$), it is known emperically~\cite{Harari:1966mu}
\begin{equation}
	\lim_{x\rightarrow 0} F_2^{p-n} \propto \sqrt{x}\
	\Rightarrow \
	\lim_{x\rightarrow 0} \textrm{Im}\, t_1^{p-n} \propto \sqrt{\nu}\, ,
\end{equation}
such that in the unsubtracted dispersion integral for $t_1^{p-n}$, the contour at infinity does not vanish but in fact diverges, see Fig.~\ref{fig:dispersion}.
One is lead to conclude that the unknown subtraction function can not be evaded and the evaluation of $\d M^\g$ in Ref.~\cite{Gasser:1974wd} is not correct.

Using either representation for $T_{\mu\nu}$, Eqs.~(\ref{eq:Tmunu}) or (\ref{eq:tmunu}), performing a once-subtracted dispersion integral at $\nu=0$, one arrives at the same answer
\begin{equation}\label{eq:dMg_terms}
 \d M^\gamma = \d M^{el} + \d M^{inel} + \d M^{sub} + \d \tilde{M}^{ct}\, .
\end{equation}
The renormalization is detailed in Ref.~\cite{Collins:1978hi}.
The operator product expansion (OPE) was used to connect the ultraviolet behavior of the integrand of Eq.~(\ref{eq:Cottingham}) with operators in the QCD+QED Lagrangian.
After canceling the log-divergence, there is a residual finite contribution, $\d \tilde{M}^{ct}$, which can be estimated using arguments similar to Naive Dimensional Analysis~\cite{Manohar:1983md};
\begin{equation}
 \d \tilde{M}^{ct}  \simeq -\frac{3\a}{4\pi} \s_{\pi N} \ln \left( \frac{\L_1^2}{\L_0^2} \right)
 	\frac{3\hat{m} -5\d}{9\hat{m}} 
	\frac{ \langle p | \bar{u} u -\bar{d}d | p \rangle}{\langle p | \bar{u} u +\bar{d}d | p \rangle}
\end{equation}
where $\s_{\pi N} =\hat{m} \langle p | \bar{u}u + \bar{d} d | p \rangle / 2M \sim 45$~MeV and the quark masses are given by $2\hat{m} = m_u + m_d$ and $2\d = m_d - m_u$.
Taking $\L_0^2 = 2 \textrm{ GeV}^2$ and $\L_1^2 = 100 \textrm{ GeV}^2$, one can estimate $|\d \tilde{M}^{ct}| < 0.02$~MeV.

The most difficult contribution to determine precisely is that from the unknown subtraction function% 
%footnote
\footnote{The same unknown subtraction function contributes to (muonic) hydrogen energy levels~\cite{Pachucki:1996zza,Carlson:2011dz,Birse:2012eb,Miller:2012ne}.} 
\begin{equation}
	\d M^{sub} = -\frac{3\a}{16\pi M} \int_0^{\L^2} dQ^2 T_1(0,Q^2)\, .
\end{equation}
The OPE constrains the high $Q^2$ behavior, $\lim_{Q^2\rightarrow \infty} T_1(0,Q^2) \propto 1 / Q^2$ and low-energy effective theory constrains the low $Q^2$ behavior~\cite{Pachucki:1996zza}
\begin{equation}\label{eq:T1lowQsq}
	\lim_{Q^2\rightarrow 0} T_1(0,Q^2) = 2\k (2+\k)
-Q^2 \bigg\{
		\frac{2}{3} \left[ (1+\k)^2 r_M^2 - r_E^2 \right]
		+\frac{\k}{M^2}
	-2 M  \frac{\b_M}{\a}
	\bigg\}
	+\mc{O}(Q^4)\, ,
\end{equation}
where $\k$ is the anomalous magnetic moment, $r_{E(M)}$ is the electric (magnetic) charge-radius and $\b_M$ is the magnetic polarizbility.
To connect the high and low $Q^2$ behavior, one must resort to modeling.
All but the last term of Eq.~(\ref{eq:T1lowQsq}) are recognized as the leading terms of elastic form factors, suggesting a resummation.
While one cannot derive the resummed formula from first principles, it can be computed from the Born graphs of nucleon-Compton scattering with the full elastic form factors inserted at the vertices~\cite{Carlson:2011dz}.
The low $Q^2$ limit of $T_1(0,Q^2)$ is now known to $\mc{O}(Q^4)$ from heavy baryon $\chi$PT and the expansion of the resummed Born contributions agrees precisely through this order~\cite{Birse:2012eb}.
Using the resummed elastic terms, one arrives at%
% FOOTNOTE
\footnote{We thank J.~McGovern for providing the isovector formula to $\mc{O}(Q^4)$ from Ref.~\cite{Birse:2012eb}.}
\begin{equation}
T_1^{p-n}(0,Q^2) \simeq 
	2G_M^2(Q^2) -2F_1^2(Q^2) 
	+\frac{2M \beta_M^{p-n}}{\a} Q^2 \left(
		1 - \frac{Q^2}{M_{\b^{p-n}}^2} \right) \, ,
	%+\frac{2}{3} \frac{g_A^2 \mu_s}{(4\pi F_\pi m_\pi)^2} Q^4\, ,
\end{equation}
with $M_{\b^{p-n}}^2 = \frac{-\b_M^{p-n}}{\a} \frac{3M(4\pi F_\pi m_\pi)^2}{g_A^2 \mu_s}$, $F_\pi \simeq 92$~MeV and the isoscalar magnetic moment $\mu_s \simeq 0.88$.
A recent review of low-energy Compton scattering yields the value $\b_M^{p-n} = -1.0 \pm 1.0 \times 10^{-4} \textrm{ fm}^3$~\cite{Griesshammer:2012we}.
If the two inelastic terms were of opposite sign, we could re-sum them to a dipole form factor as in Ref.~\cite{Birse:2012eb}.  However, the large uncertainty in even the sign of $\b_M^{p-n}$ makes this impractical to implement in a numerical evaluation.
To be conservative, we can simply multiply the leading inelastic contribution by a standard dipole form factor such that the overall contribution obeys the high $Q^2$ scaling.
This leaves us with a natural separation of the subtraction term
\begin{equation}\label{eq:sub_el_inel}
\d M^{sub}_{el} \simeq \frac{-3 \a}{8M\pi} \int_0^{\L^2} {\hskip-0.6em} dQ^2 
	\left[ G_M^2(Q^2) - F_1^2(Q^2) \right],
\quad
\d M^{sub}_{inel} \simeq \frac{-3 \b_M^{p-n}}{8\pi} \int_0^{\L^2} {\hskip-0.6em} dQ^2 
	Q^2 \left( \frac{m_0^2}{m_0^2 + Q^2} \right)^2 {\hskip-0.4em}.
\end{equation}

%
%%		Evaluation
%%%
%%%%
\section{Evaluation of $\d M^\g$}
Details of the numerical evaluation can be found in Ref.~\cite{WalkerLoud:2012bg}.  
The elastic and inelastic contributions are given by the expressions
\begin{equation}
\d M^{el} = \frac{\a}{4\pi M} \int_{0}^{\L_0^2} 
	%{\hskip-0.6em}dQ^2
	\frac{dQ^2}{\sqrt{\tel}}
	\bigg\{ 
	\frac{3\sqrt{\tel}G_M^2}{2(1+\tel)}\, 
	+
	\frac{\left[ G_E^2 -2\tel\, G_M^2 \right]}{1+\tel}
	\bigg[(1+\tel)^{3/2} -\tel^{3/2} -\frac{3}{2}\sqrt{\tel} \bigg]
	\bigg\}\, ,
\end{equation}
\begin{eqnarray}
\d M^{inel} = \frac{\a}{4\pi M} \int_{0}^{\L_0^2} \frac{dQ^2}{Q} \int_{W_{th}^2}^\infty 
	d W^2 &&
	\bigg\{
		\frac{3F_1(\nu,Q^2)}{M} \bigg[ \frac{\t^{3/2} -\t\sqrt{1+\t} +\sqrt{\tau}/2 }{\t} \bigg]
\nonumber\\ &&\qquad
		+\frac{F_2(\nu,Q^2)}{\nu} \bigg[ (1+\t)^{3/2} -\t^{3/2} -\frac{3}{2}\sqrt{\t} \bigg]
	\bigg\}\, ,
\end{eqnarray}
where $\tel = Q^2/4M^2$, $\t = \nu^2/Q^2$, $W^2 = M^2 + 2M\nu -Q^2$ and $W_{th}^2 = (M + m_\pi)^2$.
The value of $\L_0^2 = 2\textrm{ GeV}^2$ was used.
The elastic terms are evaluated with the Kelly parameterization of the elastic form factors~\cite{Kelly:2004hm} and the inelastic contributions are evaluated from parameterizations of the resonance~\cite{Bosted:2007xd} and scaling regions~\cite{Capella:1994cr}.
The subtraction terms are estimated from Eq.~(\ref{eq:sub_el_inel}), conservatively taking $m_0^2 = 0.71\textrm{ GeV}^2$.
For the isovector self-energy, we find
\begin{eqnarray}
\d M^{el} = 1.39(02) \textrm{ MeV}\, , \quad \d M^{sub}_{el} = -0.62(02) \textrm{ MeV}\, ,
\nonumber\\
\d M^{inel} = 0.057(16) \textrm{ MeV}\, , \quad \d M^{sub}_{inel} = +0.47(47) \textrm{ MeV}\, ,
\end{eqnarray}
for the total
\begin{equation}\label{eq:dMg_final}
\d M^\g = 1.30(03)(47) \textrm{ MeV}\, .
\end{equation}

%
%%		Conclusion
%%%
%%%%
\section{Concluding Remarks}
A precise determination of the electromagnetic self-energy contribution to $M_p - M_n$ remains an outstanding theoretical challenge.  We have updated the Cottingham determination of $\d M^\g$ using modern theoretical knowledge and experimental determinations of the nucleon structure functions, Eq.~(\ref{eq:dMg_final}).
In the process, a technical oversight in the old determination was uncovered invalidating the result.
The limiting factor in a precise determination comes from modeling the behavior of the unknown subtraction function between the low and high $Q^2$ limits where it is known.
Presently, the large uncertainty can be traced directly to the poorly determined isovector magnetic polarizability.
Once this quantity is precisely known, an improved understanding of $T_1^{p-n}(0,Q^2)$ will be required to make progress.

%\newpage

\end{document}